\documentclass[a4paper]{jpconf}

\usepackage{enumitem}
\usepackage{graphicx}
\usepackage{caption}

\newcommand{\Osq}{O\textsuperscript{2}}

\begin{document}
\title{The ALICE \Osq{} common driver for the C-RORC and CRU read-out cards}
\author{Boeschoten P and Costa F for the ALICE collaboration}
\ead{pascal.boeschoten@cern.ch, filippo.costa@cern.ch}


\begin{abstract}
ALICE (A Large Ion Collider Experiment) is the heavy-ion detector designed to study the strongly interacting state of
matter realized in relativistic heavy-ion collisions at the CERN Large Hadron Collider (LHC). A major upgrade of the
experiment is planned during the 2019-2020 long shutdown. In order to cope with a data rate 100 times higher than during
LHC Run 1 and with the continuous read-out of the Time Projection Chamber (TPC), it is necessary to upgrade the Online
and Offline Computing to a new common system called \Osq{}. The \Osq{} read-out chain will use commodity x86 Linux servers
equipped with custom PCIe FPGA-based read-out cards. This paper discusses the driver architecture for the cards that will
be used in \Osq{}: the PCIe v2 x8, Xilinx Virtex 6 based C-RORC (Common Readout Receiver Card) and the PCIe v3 x16, Intel
Arria 10 based CRU (Common Readout Unit). Access to the PCIe cards is provided via three layers of software. Firstly,
the low-level PCIe (PCI Express) layer responsible for the userspace interface for low-level operations such as memory
mapping the PCIe BAR (Base Address Registers) and creating scatter-gather lists, which is provided by the PDA (Portable
Driver Architecture) library developed by the Frankfurt Institute for Advanced Studies (FIAS). Above that sits our
userspace driver which implements synchronization, controls the read-out card -- e.g. resetting and configuring the card,
providing it with bus addresses to transfer data to and checking for data arrival -- and presents a uniform, high-level
C++ interface that abstracts over the differences between the C-RORC and CRU. This interface -- of which direct usage is
principally intended for high-performance read-out processes -- allows users to configure and use the various aspects of
the read-out cards, such as configuration, DMA transfers and commands to the front-end. The top layer consists of a
Python wrapper and command-line utilities that are provided to facilitate scripting and executing tasks from a shell,
such as card resetting; performing benchmarks; reading or writing registers; and running test suites. Additionally, the
paper presents the results of benchmarks in various test environments. Finally, we present our plans for future
development, testing and integration.
\end{abstract}

\section{Introduction}
ALICE\cite{ALICE} (A Large Ion Collider Experiment) is an experiment at the CERN LHC\cite{LHC} (Large Hadron Collider)
studying the Quark-Gluon Plasma -- a state of matter which existed shortly after the Big Bang -- exploiting heavy-ion
collisions.
In 2019-2020, the LHC will undergo its second long shutdown (LS2) to prepare for Run 3, after which it will run at
significantly higher luminosity.
During the shutdown ALICE will upgrade its detectors and software systems to handle the higher data rate of over 3.4
TB/s. The new computing software suite called \Osq{}\cite{Osquare} will do both offline and online processing. Part of
this suite is the ReadoutCard\cite{ReadoutCard} library containing the driver for the two read-out cards used during Run 3.
In this paper, we will discuss the design and capabilities of the ReadoutCard library, the supported cards, and present
benchmark results.

\section{Cards}
\begin{figure}
\centering
\begin{minipage}{.5\textwidth}
  \centering
  \includegraphics[width=.9\linewidth]{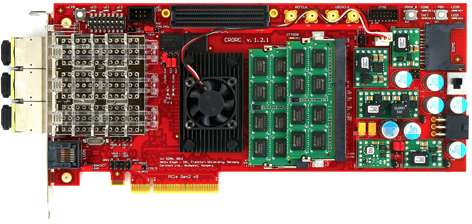}
  \captionof{figure}{C-RORC}
  \label{fig:c-rorc}
\end{minipage}%
\begin{minipage}{.5\textwidth}
  \centering
  \includegraphics[width=.9\linewidth]{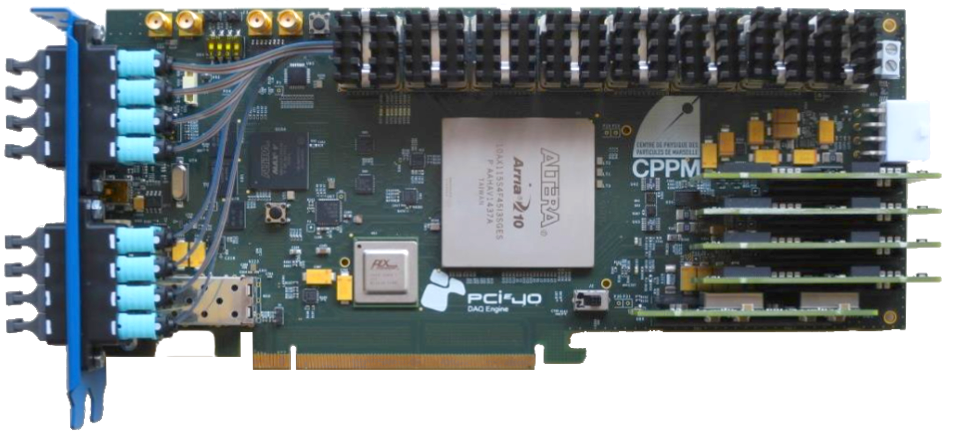}
  \captionof{figure}{CRU}
  \label{fig:cru}
\end{minipage}
\end{figure}

The cards supported by the ReadoutCard library are the C-RORC\cite{C-RORC} (Common Readout and Receiver Card) seen in
figure \ref{fig:c-rorc}, and the CRU (Common Readout Unit) shown in figure \ref{fig:cru}. Both are custom-built
FPGA-based PCIe cards.
The C-RORC, is based on a Xilinx Virtex 6 FPGA. It supports up to PCIe v2x8, with the current firmware implementing v2x4
only (as the additional bandwidth was unnecessary). The card can take up to 12 optical links and distribute the data
over 6 DMA channels to the host PC at a total maximum throughput of 1.6 GB/s. It is currently in use in ALICE for Run 2
and 25 units will be reused for some of the subdetectors during Run 3.

The CRU is based on the Intel Arria 10 FPGA, using PCIe v3x16, supporting 48 optical links, and 2 DMA channels with a
total throughput of ~27 GB/s measured using the current firmware and software. It will be the main read-out card for
ALICE during Run 3 with 461 units to be used.

\section{Software}
The software stack is subdivided into three layers:
\begin{itemize}
  \item Low-level PCIe system handling, provided by the PDA\cite{Pda} library
  \item A userspace driver with C++ interface for high-performance read-out, providing the core functionality
  \item A Python interface and command-line utilities that use the C++ interface
\end{itemize}
The last two are referred to as the ReadoutCard library, which is part of the \Osq{}
software suite.

\subsection{PDA}
PDA -- the result of extensive work described in D. Eschweiler's thesis\cite{Eschweiler} -- consists of a UIO (Userspace
IO) kernel module that handles the creation of scatter-gather lists for DMA buffers, and a userspace C library that
interfaces with the kernel module and provides additional utilities such as:
\begin{itemize}
  \item PCI device enumeration.
  \item Registering DMA memory targets with the IOMMU (Input Output Memory Mapping Unit). This component can take
        multiple separate memory regions and present them as a contiguous virtual memory space to the PCI device. This
        is analogous to how the regular MMU (Memory Mapping Unit) works, presenting scattered physical memory regions as
        a single virtual space to processes running on the CPU. Both the IOMMU and MMU also provide protection against
        invalid memory accesses.
  \item Generation of scatter-gather lists, which combines regions of memory into a single list for the card's DMA
        engine. This is essential if an IOMMU is not enabled or not present.
  \item Memory mapping of the device's BAR (Base Address Register), a range of device registers memory-mapped to a
        virtual address space so they can accessed with regular memory reads and writes.
\end{itemize}

\subsection{Userspace driver}
The driver wraps the PDA functions in C++ classes, and builds upon the tools they provide to configure and
control the cards.
For each card, the ReadoutCard's C++ interface exposes to the user a single instance -- system-wide -- of a
DmaChannelInterface implementation. The class instantiated is specific to the type of the card, and it operates the
card's machinery. Read-out processes use it for high-performance data taking. It provides a queue-style interface, to
push DMA target addresses and pop them when the transfer is complete. To instantiate this class, the user must provide a
DMA buffer, which will be registered internally with PDA to create the scatter-gather list.
The buffer can be any type  of shared memory if the IOMMU is enabled, since the IOMMU can present the memory as a single
contiguous region to the card. However, if the IOMMU is not available, it is essential to use hugepage-backed shared
memory to ensure the buffer meets the required physical contiguousness for the cards' DMA engines. This is enforced by
the driver through a lookup in the process's memory mappings.

In addition, the interface can provide multiple instances of BarInterface classes per card.
These are essentially wrappers around the BAR, providing simple read and write access with bounds checking.
They may also prohibit reads and/or writes to certain address ranges under certain conditions to prevent malfunctions.

The library uses ``Boost Exception'' for its error handling. When these exceptions are raised,
arbitrary data can be added to them while they are caught and rethrown, bubbling up to the final catch site with
information about possible causes and potential fixes for the error. This is useful for issues where the program itself
cannot easily deduce the cause of the error. In these cases, the rich diagnostic messages generated from the exception
allow users to potentially fix the issues themselves quickly, without referring to in-depth documentation or contacting
developers.

\subsection{Python interface}
The Python interface is a wrapper around a subset of the C++ interface: the BarInterface. It aims to provide easy access
for scripting purposes, and other situations where flexibility and ease of use is required. For example, for development
of the card's firmware or detector. We expect it will be mainly used by card firmware developers, and detector teams
testing configuration which are passed through the card to the detector front-ends.

\subsection{Command-line utilities}
The module contains several utility programs to assist with development, debugging and administration. Every program
will display detailed information and a usage example when ran with the --help option. Most programs will also provide
more detailed output when given the --verbose option.
The available programs are:
\begin{description}[align=left,labelwidth=3cm]
  \item[roc-bench-dma] DMA throughput and stress-testing benchmarks.
  \item[roc-flash] Flashes firmware from a file onto the card.
  \item[roc-flash-read] Reads from the card's flash memory.
  \item[roc-list-cards] Lists the read-out cards present on the system, along with their type, PCI address, vendor ID,
    device ID, serial number, and firmware version.
  \item[roc-reg-read] Reads from registers on a card's BAR.
  \item[roc-reg-write] Writes to registers on a card's BAR.
  \item[roc-reg-reset] Resets a card channel.
\end{description}

\section{Buffer layout \& data format}

The buffer of a DMA channel is referred to as the channel buffer. In real-world running conditions, these will be large --
in the order of tens of gigabytes -- to accommodate latency in the processing chain. It is allocated by the read-out task
and registered with the driver before starting a data taking run.
The read-out is then responsible for subdividing the channel buffer into ``superpages''. These are the logical units of
transfer for the driver, and will be filled with multiple data pages by the cards. They are passed to the driver as a
memory region within the DMA buffer, ranging in size from 32 kilobytes to gigabytes.
The C-RORC and CRU are cards with very different interfaces, so the driver will then take care of the transfer in a
card-specific way. For the older C-RORC, superpages are filled up with multiple transfers.
The CRU, whose firmware was developed in close cooperation with the driver, natively supports superpages: it is given a
pointer and size, and simply fills the entire superpage with data.

The in-memory data format consists of 8 kilobyte DMA pages. They each contain a header, the payload, a trailer, and
padding in case the data does not fill up the DMA page.
This rigid structure aids recovery or inspection of data in case of transfer errors.
In addition, the driver will scan the superpage and generate a table of indices corresponding to DMA pages that start
at ``heartbeats'', special DMA pages that indicate synchronization intervals or conditions. This lookup table speeds
up handling and processing along the read-out chain.

\section{Benchmark}
Throughput was measured on a Dell R730 server (2x Intel E5-2630 v3 CPUs, 2133 MHz DDR4 RAM) fitted with two CRU cards,
using various superpage sizes, recording individual DMA channel throughputs and summing them to arrive at a total
throughput for both cards. The operating system used was CERN CentOS 7.
The required throughput specified for the \Osq{} system is 5.625 GB/s. Additionally, we set a development target for
6.9 GB/s to allow for headroom.

Data was gathered using a script\cite{MultiBench} that simultaneously starts instances of the roc-bench-dma program for
each DMA channel. The cards were configured to use an internal data generator during the minute-long runs.
To optimize the throughput on this multi-socket server, it was necessary to bind the DMA processes' threads and memory
allocation to the CPU whose PCIe lanes are directly connected to the corresponding card. To achieve this, the script
uses the ``numactl'' utility to launch the processes.

\begin{figure}
  \centering
  \includegraphics[width=.65\linewidth]{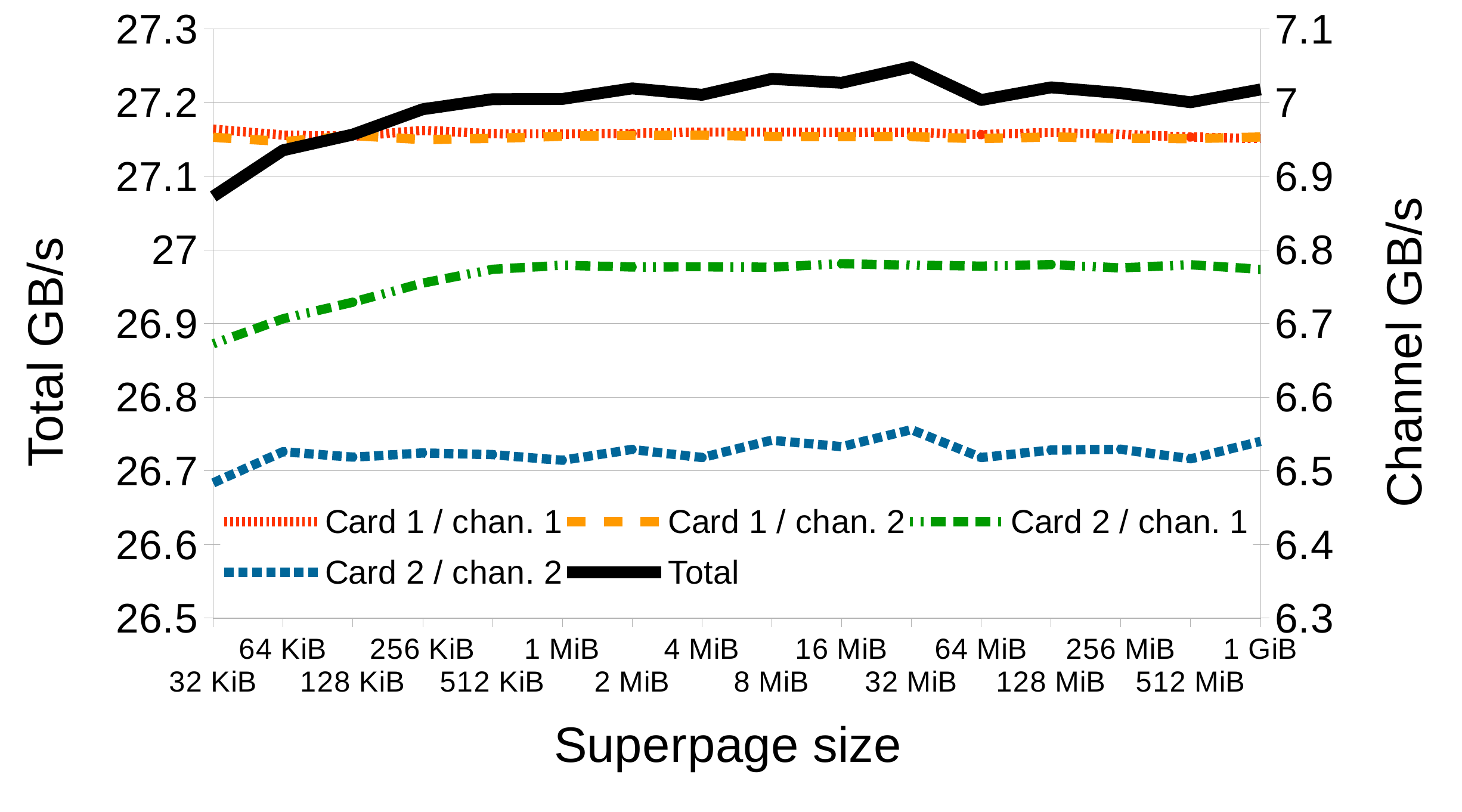}
  \captionof{figure}{CRU benchmark results}
  \label{fig:cru-bench}
\end{figure}

Figure \ref{fig:cru-bench} shows the results of the benchmark, with total and channel throughputs displayed as a
function of superpage size. Little variance in total throughput is shown, even between the smallest supported superpage
sizes and very large ones. For all channels, the minimum required throughput was exceeded. Both channels of card 1 were
consistently performing similarly, exceeding the target throughput. Both of card 2's channels did not reach the target
throughput. Card 2 channel 1 showed a steady increase up to 1 MiB superpages, then remained steady. Card 2 channel 2
performed inconsistently and is responsible for what jitter there is in the total throughput. Note that the DMA engines
on all of these cards and channels are identical. More work is being done to understand the throughput differences
between channels.

\section{Conclusions}
In this paper, we have described a userspace driver for high-performance read-out built upon PDA, its supporting
utilities, the cards and data structures involved, and presented benchmark results demonstrating the driver's
capabilities. Our next steps are:
\begin{itemize}
  \item Testing of new server models coming from different companies to identify the best candidate for the \Osq{} farm.
  \item Investigate throughput differences between individual channels when running with multiple cards.
  \item Adapt the driver to CRU firmware changes, following experiment requirements.
  \item Finalize documentation and software interface, so detector teams can use the \Osq{} software and connect their
        own modules to the \Osq{} framework.
\end{itemize}


\end{document}